# Multiplex Media Attention and Disregard Network among 129 Countries

Haewoon Kwak
Qatar Computing Research Institute
Hamad Bin Khalifa University
Doha, Qatar
Email: haewoon@acm.org

Jisun An
Qatar Computing Research Institute
Hamad Bin Khalifa University
Doha, Qatar
Email: jisun.an@acm.org

*Abstract*—We built a multiplex media attention and disregard network (MADN) among 129 countries over 212 days. By characterizing the MADN from multiple levels, we found that it is formed primarily by skewed, hierarchical, and asymmetric relationships. Also, we found strong evidence that our news world is becoming a "global village." However, at the same time, unique attention blocks of the Middle East and North Africa (MENA) region, as well as Russia and its neighbors, still exist.

## I. INTRODUCTION

The world is rapidly globalized with evolving technology in communication and commutation [1]. In this "global village," products, capital, technology, and information are freely moving around the world, and we can easily access and consume those products and information from abroad. However, one major constraint still remains–our understanding of the world is dominantly shaped by the media [2]. What we see, read, and hear about other countries is a result of the gatekeeping by the journalists and of the social, economic, and political relationships across countries [3]. Thus, international news coverage plays a central role as a medium that represents the people's shaping images of other countries in understanding the progress of globalization.

Researchers have studied a pairwise international news coverage. Typically, the studies follow two approaches. First, they have applied the theory of news values, such as unexpectedness, proximity, conflict, discrepancy, and prominence, to study why some countries more likely than other to be covered [4], [5], [6], [7]. The other approach focused on different systematic factors of international relationships such as trade, territorial size, cultural ties, communication resources, and physical distance in influencing the news coverage [3], [8], [9], [10].

Previous studies altogether have established ground theories in understanding international news coverage by analyzing pairwise country relationships. However, the impact of the international news flow is beyond the pairwise relationship. Since a certain set of countries are tightly connected in terms of media coverage, a story published in one country can easily spread to other close countries. If the story has a negative tone, such as fear, the "fear" about a country can spread from one country to another not because of the news value or any systematic factors, but because of the international news flow.

Thus, it seems natural to analyze the media attention beyond the view of pairwise relationships.

This study aims to explore the nature of international media attention across 129 countries around the world. To this end, we build a multiplex media attention and disregard network (MADN). As its name says, adding to the attention, we also model disregard ("ignorance with intention") for news media, which is relatively unexplored. We characterize the MADN at multiple scales, from micro-scale to meso- and macro-scale, leading us to unveil its complex structure.

The contribution of this work is three-fold. First, we introduce a new dimension of media attention: disregarding relationships. Previous work had focused on what media pay attention to, but not what media disregard. By adding this new dimension to the news world, we enrich the study of international relationships. Secondly, we build the MADN to understand international media attention. Unlike previous work, which focused only on pairwise relationships, our work is the first to reveal the network structure of media attention. Our results reveal that the MADN is formed primarily by the skewed, hierarchical, and asymmetric relationships. Lastly, we find strong evidence that our news world is becoming a global village. However, at the same time, unique attention blocks of the Middle East and North Africa (MENA) region, as well as Russia and its neighbors, still exist.

## II. RELATED WORK

**News selection.** Not all the events happening in the world are covered by news media. The news selection process is often guided by an understanding of news values, a set of criteria to determine the newsworthiness of an event. Galtung and Ruge ([4]) firstly presented twelve factors, including frequency, unexpectedness, and reference to the elite nations/people, that they intuitively identified as being important in the selection of news. Nearly 40 years after Galtung's work, Harcup and O'Neil proposed a new set of news values: The Power Elite, Celebrity, Entertainment, Surprise, Bad News, Good News, Magnitude, Relevance, Follow-up, and Newspaper Agenda [6]. While the previous studies successfully developed the taxonomy of news values, media attention among countries remains unexplored using a large-scale data set.

**News and international relations.** International news coverage is known to have a different news selection mechanism from domestic news coverage–it is a result of both news values and the current practice of news gathering around the world [3]. Scholars have focused on understanding the current global divides and power structure through news coverage. Some studies claimed that Western news media reinforce a particular kind of world order, mainly a Euro-American-centered one [11], [12].

In understanding factors relating to foreign news coverage, based on a small amount of data, studies showed inconsistent results due to cultural, regional, or political differences [8]. The "Foreign News" study dataset [13], containing news media from 46 countries, has been used in several studies to discover patterns of global news flow [10] where geographical and economic proximity play a major role in the news selection process. Recently, using the GDELT dataset, Kwak and An revealed a strong regionalism and a power structure among countries [9]. In this work, we examine international news coverage to reveal the network structures of media attention.

**Summary.** Media attention has been studied over decades from multiple perspectives including news selection and international relations. Most studies have been conducted with limitations imposed by studying a small handful of countries, news articles, or topics. We are privileged in the digital era by the emergence of advanced machine learning techniques so we can overcome these limitations and provide a data-driven analysis of media attention for 129 countries over 212 days.

## III. Building Multiplex Network

We build a multiplex media attention and disregard network (MADN) among 129 countries. The multiplex network is one of the specific realizations of multi-layer networks, having a constraint that all the layers have the same set of nodes. We use $N^M$ to refer to the MADN, $N^A$ to the attention network, and $N^D$ to the disregard network.

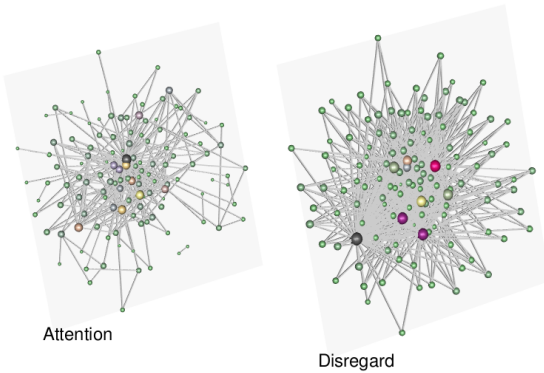

Fig. 1. MADN composed of attention and disregard networks

We collect news data from one of the biggest news aggregation sites, Unfiltered News, based on the Google News database run by Alphabet from 7 March 2016 to 9 October 2016 (212 days in total, denoted by T). The granularity of the

| | $N$ | $L$ | $\langle k \rangle$ | $CC$ | $a$ | $SCC(\%)$ | $\langle d \rangle$ |
|---|---|---|---|---|---|---|---|
| $N^A$ | 129 | 3,058 | 23.7 | 0.134 | 0.077 | 94.6 | 2.184 |
| $N^D$ | 129 | 3,989 | 30.9 | 0.133 | -0.116 | 57.4 | 1.613 |

data is not article-level but aggregated-level. Each record $[t, c_i, e, n(e)_{c_i}^t]$ consists of four values: 1) $t$: date, 2) $c_i$: the origin country of the news media, 3) $e$: the entity mentioned in the news media of $c_i$, and 4) $n(e)_{c_i}^t$: the normalized number of the news articles mentioning $e$ in $c_i$ at $t$. A full list of entities is available in Google Knowledge Graph, which is a knowledge base like Wikidata, and translated by a cross-lingual database.

Using top $k$ records with the highest $n(e)_{c_i}^t$ of each country per day, we build a daily $N_t^A$ where $t \in T$ in the following ways: 1) if any record at $t$ contains a country $c$ as an origin country of news media, then the country $c$ becomes a node in $N_t^A$; and 2) if the news media of an origin country $c_i$ mentions another country $c_j$ in a certain record on $t$, we create a link from $c_i$ to $c_j$ in $N_t^A$. Note that $N_t^A$ is a directed and unweighted network, and self-loops are eliminated.

Once we have $N_t^A$ for all $t \in T$, we superimpose them to build $N^A$: 1) if a node or a link exists in any $N_t^A$, they exist in $N^A$ as well; and 2) a weight of the link from $c_i$ to $c_j$ in $N^A$ is the number of $N_t^A$ containing the corresponding link where $t \in T$ (i.e. the number of days $c_i$ mentioning $c_j$).

The main challenge in building $N^D$ is to differentiate disregard (i.e., editors choose not to report a certain incident) from ignorance (i.e., journalists are not aware of the incident). The workflow of newsroom using the story alert from international news wire services allows us to reasonably assume that a widely reported incident is highly likely to be shared through those services. In this case, we can say that such an incident is disregarded if one country does not report it.

More formally, Unfiltered News computes the level of disregard of news media in country $c$ toward an entity $e$ on a certain date $t$ as: $\Delta(e)_c^t = \frac{\sum_{c \in C} n(e)_c^t}{n(e)_{c_i}^t + 0.1}$ , where $\sum_{c \in C} n(e)_c^t$ is the number of news articles mentioning the entity $e$ on a date $t$ across all the countries C, and 0.1 is the constant to avoid the division by zero. The higher $\Delta(e)_c^t$ means that the entity $e$ is more disregarded by news media in the country $c$ on the date $t$. We then are able to build a daily $N_t^D$ using the top $k$ entities with highest $\Delta(e)_c^t$ within each $c$. We omit the detailed process to build $N_t^D$ and $N^D$ as it is conceptually the same as the process to build $N_t^A$ and $N^A$. In this work, we set $k$=10 in building both $N_t^A$ and $N_t^D$. We conducted all our experiments with different $k$ (10 to 90) and found that overall trend stays the same with some variations in numbers. Figure 1 shows our MADN composed of attention and disregard networks.

## IV. Structures of MADN in Multi-level

### A. Topological Characteristics of Networks

To show the basic characteristics of our MADN, we first look into the overall property of both networks, $N^A$ and $N^D$,

separately. Table I shows the number of nodes $N$, that of links $L$, mean degree $\langle k \rangle$, mean clustering coefficient $CC$, assortativity coefficient $a$, percentage of nodes in the strongest connected component $SCC(\%)$, and mean shortest path length of the strongest connected component $\langle d \rangle$ of $\text{N}^\text{A}$ and $\text{N}^\text{D}$. In short, both networks commonly have: 1) low clustering coefficient, 2) almost neutral assortativity, and 3) short average path length.

To fully understand the structural characteristics of the MADN, we explore it from multiple views including microscale to mesoscale, and macroscale.

### B. At Node Level: Country Centrality

We begin with a node level property, degree centrality. It shows the most frequently covered or disregarded countries (from indegree) and the countries that have the most or least interest in foreign issues (from outdegree).

| Rank | $\text{N}^\text{A}$ | | | $\text{N}^\text{D}$ | |
| --- | --- | --- | --- | --- | --- |
| | $k_{out}$ | $k_{in}$ | PageRank | $k_{out}$ | $k_{in}$ |
| 1 | French Polynesia | UK | USA | Jordan | Ukraine |
| 2 | Finland | USA | Syria | UK | Yemen |
| 3 | Switzerland | France | Turkey | Israel | Turkey |
| 4 | Liechtenstein | Turkey | Russia | Kyrgyzstan | Egypt |
| 5 | Oman | Belgium | UK | Zimbabwe | China |
| 6 | Mauritania | Italy | Saudi Arabia | Yemen | Saudi Arabia |
| 7 | El Salvador | Syria | Ukraine | Russia | Russia |
| 8 | Bahrain | Russia | France | Russia | Iran |
| 9 | Malta | Panama | Egypt | Pakistan | Syria |
| 10 | Costa Rica | China | Belgium | Cameroon | Germany |

Table II shows that most countries with a high outdegree are relatively small countries. The higher outdegree means that they actively report foreign issues across the world. One possible explanation is that it is cost-effective to reformat the news produced by international news agencies (e.g. Reuters or Associated Press) and distribute it to domestic readers. This tendency could be accelerated when the market is small, and foreign news coverage becomes significant in those countries.

In $\text{N}^\text{A}$, the UK is the top country by indegree (126), but the USA has the highest PageRank. While indegree reflects how many countries have paid attention to the corresponding country, PageRank considers the importance of each country mentioning another country. This once again confirms that "the USA is placed in the brightest spotlight on the stage of the news world [3]." Along with the UK and the USA, some differences are noticeable between top countries by indegree and PageRank, implying that the centrality measure in $\text{N}^\text{A}$ should be carefully chosen for different purposes.

The median of indegree in $\text{N}^\text{A}$ is 10, and that of outgoing links is 23. However, their distributions are highly skewed. We fitted the indegree distribution in the exponential distribution ($P(k_{in}) \propto e^{-k/0.0415}$ with $k_{min} = 5$), while the outdegree distribution has a tail whose very small proportion fits in the power-law ($k_{min} = 37$).

Among 129 countries on $\text{N}^\text{D}$, 55 countries do not have any incoming link. The top five countries with the highest indegree are Ukraine, Yemen, Turkey, Egypt, and China. Interestingly, we find some countries that are in the top ranks of indegree in $\text{N}^\text{D}$ and $\text{N}^\text{A}$ simultaneously. This implies that a country has different news values for different countries; even though one country is highly reported due to a certain incident, some countries do not take its news value as high as other countries do.

### C. At Dyadic Level: Media Attention Bias

The variations in news value of different countries form links differently in $\text{N}^\text{A}$ and $\text{N}^\text{D}$ and thus lead to different centrality in each network. How, then, do the links interplay between the two networks? To answer this question, we compare the weights of the links between $\text{N}^\text{A}$ and $\text{N}^\text{D}$.

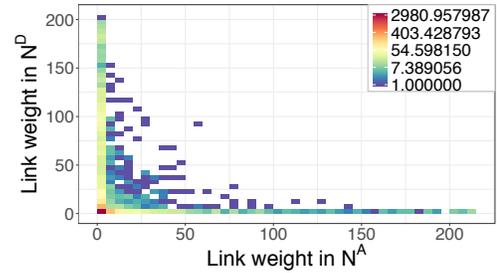

Fig. 2. Weights of links in $\text{N}^\text{A}$ and $\text{N}^\text{D}$

Links appearing in both $\text{N}^\text{A}$ and $\text{N}^\text{D}$ mean that sometimes one country pays more attention to another country and at other times disregards the same country. As the difference in the weights from both networks gets higher, the media attention becomes less flexible and more biased.

Figure 2 shows how each link has a different weight in $\text{N}^\text{A}$ and $\text{N}^\text{D}$. The size of each rectangle is 5×5, and its color represents the number of links that have corresponding weights. In the figure, most of the colored boxes are closer to either the $x$- or $y$-axis. This indicates a link with a higher weight in $\text{N}^\text{A}$ is more likely to have a smaller weight (even zero) in $\text{N}^\text{D}$ and vice versa. In other words, news media in one country $c_i$ that often mention another country $c_j$ (i.e., high weight of $\overrightarrow{c_ic_j}$ in $\text{N}^\text{A}$) do not disregard $c_j$ when news media in other countries mention $c_j$ (i.e., low weight of $\overrightarrow{c_ic_j}$ in $\text{N}^\text{D}$). Likewise, news media in one country $c_i$ that often disregard another country $c_j$ (i.e., high weight of $\overrightarrow{c_ic_j}$ in $\text{N}^\text{D}$) rarely pay attention to the same $c_j$ (i.e., low weight of $\overrightarrow{c_ic_j}$ in $\text{N}^\text{A}$).

The result affirms the existence of media attention bias. In the next section, we examine how asymmetric the attention or disregard relationships are.

### D. At Dyadic Level: Media Attention Asymmetry

The notion of a media attention direction captures the existence of a level of symmetry between two countries. The standard measure for the level of symmetry in a given network is its link reciprocity, which is defined as the likelihood that the link in the opposite direction exists when a link from one node to another node exists. It is known that social systems generally have high reciprocity [14].

The reciprocity is very low in both networks; it is 0.156 in $N^A$ and 0.168 in $N^D$. In other words, only 15.6% of all the pairs of countries pay attention to each other. Similarly, 16.8% of all the pairs of countries disregard each other. This shows that unequal, imbalanced media attention and disregard exist between countries. A real network that has low reciprocity is the Twitter follower network (0.221) [15]. In the rest of this section, we unveil the asymmetry structure of the MADN, with a follow-up analysis.

*1) Pairwise Relationships of Media Attention:* Here we look into how a pairwise country relationship is defined in the MADN. To this end, we combine two networks, $N^A$ and $N^D$, and when the links exist in both networks, we take the stronger relationship (the higher weight).

TABLE III
Breakdown of pairwise relationships

| Type | | Relationships | Percentage |
|---|---|---|---|
| ($\overrightarrow{A}$ | $\varnothing$) | 644 | 22.9% |
| ($\overrightarrow{D}$ | $\varnothing$) | 1129 | 40.1% |
| ($\overrightarrow{A}$ | $\overleftarrow{A}$) | 315 | 11.2% |
| ($\overrightarrow{D}$ | $\overleftarrow{D}$) | 466 | 16.6% |
| ($\overrightarrow{A}$ | $\overleftarrow{D}$) | 212 | 7.5% |

Table III shows the breakdown of pairwise relationships by their attention and disregard patterns. More than half of the relationships are unidirectional (63.0%), as we expected from the low reciprocity. This clearly shows the inherent asymmetric nature of the MADN.

Of the relationships, 11.2% are of two countries exchanging their media attention. Geographically close countries tend to have this relationship; news media in neighboring countries actively report issues of each other. Also, the relationship between the UK and the USA is also in this category, showing their strong connection.

Of the relationships, 16.6% are between countries are that disregard each other. Most of them are led by coverage of China, Yemen, Saudi Arabia, or Ukraine. We also find that news media in Turkey and the United Kingdom disregard each other. This finding is particularly surprising because it is contrary to the previous finding that the volume of economic transactions is highly correlated with the foreign news coverage [3] – the UK is the second biggest importer of goods from Turkey as of 2015. This demonstrates that the notion of disregard adds a new dimension to research on foreign news coverage and brings new research opportunities.

Even though its proportion is the lowest (7.5%), the most unexpected relationship is that news media in two countries behave in opposite ways. One country pays attention to another country but this other country disregards the first country, showing imbalanced and unequal media attention. Among 212 such relationships, 29 are paying attention to the USA and 17 to the UK, placing them at the "global hubs" in the MADN. The other ends of those relationships are "local hubs," which are Germany, France, China, Saudi Arabia, and Russia. While these local hub countries are actively mentioned by news media in neighboring countries, news media in the USA or the UK probably mention them less than the other neighbors do. Thus, this leads to a hierarchical structure of the news world; the transitive hierarchy exists among a small country, a local hub, and a global hub, in that the actor in this hierarchy pays attention to upper-level actors only. This hierarchical structure is also discovered by motif analysis in Section IV-E.

*2) Taxonomy of Media Attention Relationship:* In the previous section, we learned that the MADN has an asymmetric or hierarchical structure. However, it misses one important aspect of a weighted network: not all the pairwise relationships may be significant. By considering the significance of the links, we can investigate media attention asymmetry in depth.

As $N^A$ and $N^D$ show large variations in node strength and link weight, measuring the importance of links is not straightforward. For example, the significance of the link with a weight of 10 should be different from the node with a strength of 20 and one with 100. The disparity filter proposed by Serrano et al. [16] addresses this issue and extracts significant links (called a "backbone") of a given network by considering the strength of a node and the weight of the adjacent links together.

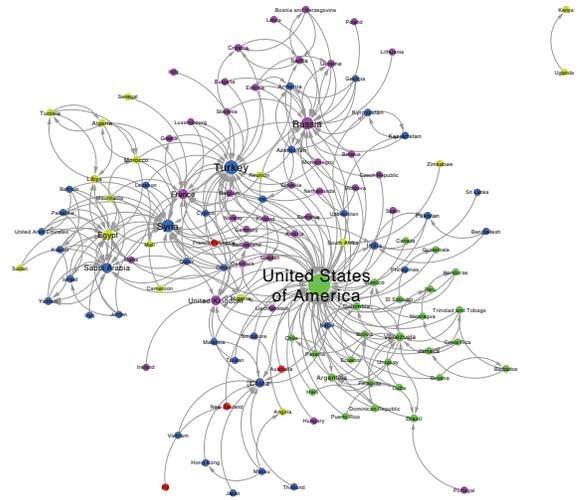

Fig. 3. [Zoomable in PDF] Backbone of $N^A$

Figure 3 shows the backbone of $N^A$. It has a complex structure where 56 countries have at least one incoming link, forming a giant (weakly) connected component. However, a couple of countries are not a part of this component (attention from Uganda to Kenya). $N^D$ has a simpler backbone structure than $N^A$, having only 10 countries with incoming links. It leads to embedded star-shaped structures in $N^D$. We omit the illustration of the backbone of $N^D$.

Using the backbone, we propose a novel method to characterize pairwise country relationships by the following criteria:

1) Is the link from $c_i$ to $c_j$ significant to $c_i$ compared to other links from $c_i$?
2) Is the link from $c_i$ to $c_j$ significant to $c_j$ compared to other links to $c_j$?

In other words, the significance of the link from one country to another is measured from the perspectives of both the sender and receiver. Then, we can think of $2 \times 2 = 4$ cases of significance for a given link.

For the sake of simplicity, we use two symbols, $\triangleright$ and $\cdot$, representing significance and non-significance, respectively. Then, every link can be represented as three symbols: 1) significance of the link to the sender, 2) direction of the link, and 3) significance of the link to the receiver. [Japan $\cdot \rightarrow \triangleright$ South Korea] means that 1) $\cdot$: this link is not significant to Japan compared to other links from Japan, 2) $\rightarrow$: this link is from Japan to Korea, and 3) $\triangleright$: this link is significant to Korea compared to other links to Korea. Considering the opposite direction of the link as well, the pairwise country relationship can be written as [Japan $\cdot \rightarrow \triangleright \cdot \leftarrow \cdot$ South Korea]. As a result, there are 10 different types of pairwise country relationships: $2^4$ combinations - 6 duplicates ( $[\cdot \rightarrow \triangleright \cdot \leftarrow \cdot]$ is the same as $[\cdot \rightarrow \triangleleft \leftarrow \cdot]$ for both ends are exchangeable).

TABLE IV
TAXONOMY OF MEDIA ATTENTION EXCHANGE

| Desc | | Count | Example $[c_i, c_j]$ |
|---|---|---|---|
| $c_i \triangleright \rightarrow \triangleright$ | $\triangleleft \leftarrow \triangleleft c_j$ | 10 | [India, Pakistan], [Russia, Ukraine] |
| $c_i \triangleright \rightarrow \triangleright$ | $\triangleleft \leftarrow \cdot c_j$ | 12 | [Cyprus, Greece], [Bangladesh, India] |
| $c_i \triangleright \rightarrow \triangleright$ | $\cdot \leftarrow \triangleleft c_j$ | 2 | [Serbia, Croatia], [Venezuela, Columbia] |
| $c_i \triangleright \rightarrow \triangleright$ | $\cdot \leftarrow \cdot c_j$ | 108 | [Armenia, Turkey], [Japan, China] |
| $c_i \triangleright \rightarrow \cdot$ | $\triangleleft \leftarrow \triangleleft c_j$ | 5 | [Kuwait, Saudi Arabia], [Ecuador, Columbia] |
| $c_i \triangleright \rightarrow \cdot$ | $\triangleleft \leftarrow \cdot c_j$ | 1 | [UK, USA] |
| $c_i \triangleright \rightarrow \cdot$ | $\cdot \leftarrow \triangleleft c_j$ | 133 | [53 countries, USA], [21 countries, Turkey] |
| $c_i \cdot \rightarrow \triangleright$ | $\triangleleft \leftarrow \cdot c_j$ | 2 | [Singapore, Malaysia], [Bahrain, Jordan] |
| $c_i \cdot \rightarrow \triangleright$ | $\cdot \leftarrow \cdot c_j$ | 65 | [Tunisia, France], [Hong Kong, UK] |
| $c_i \cdot \rightarrow \cdot$ | $\cdot \leftarrow \cdot c_j$ | 2307 | [Brazil, Mexico] |

Table IV shows the ten different types of pairwise country relationships in $N^A$. While 91.2% of country relationships are non-significant for both countries for both directions, others show the complex nature of media attention exchanged between two countries. To get a better understanding of the table, we explain the meaning of the basic building blocks in detail. First, $[c_i \triangleright \rightarrow ... c_j]$ shows the high dependence of media attention of $c_i$ on $c_j$. In other words, news media in $c_i$ more frequently cover $c_j$ than other countries. Second, $[c_i \cdot \rightarrow ... c_j]$ shows that $c_i$ more frequently covers other countries than $c_j$. Third, $[c_i ... \rightarrow \triangleright c_j]$ presents that $c_i$ covers $c_j$ more frequently than other countries do. In other words, whenever $c_j$ is covered by foreign countries, one of them is highly likely to be $c_i$. This implies that the newsworthiness of $c_j$ is higher for $c_i$ than other countries perceive. Fourth, $[c_i ... \rightarrow \cdot c_j]$ exhibits that $c_j$ is covered more frequently by other countries than $c_i$. When $c_j$ is a newsworthy country, such as the USA or the UK, this pattern emerges as $[c_i \triangleright \rightarrow \cdot c_j]$.

Then, there are some interesting patterns. $[c_i \triangleright \rightarrow \triangleright \triangleleft \leftarrow \triangleleft c_j]$ shows a strong dependency of media attention; some neighboring countries, such as Libya and Tunisia, Morocco and Algeria, or Yemen and Saudi Arabia, have this relationship. Some hub countries, such as the UK, the USA, or Saudi Arabia, are involved in $[c_i \triangleright \rightarrow \cdot ... c_j]$ relationships. Particularly, $[c_i \triangleright \rightarrow \cdot \cdot \leftarrow \cdot c_j]$ shows the asymmetry between typical countries ($c_i$) and hub countries ($c_j$), such as the USA, Turkey, or Russia.

$[c_i \triangleright \rightarrow \cdot \cdot \leftarrow \triangleleft c_j]$ shows the uniqueness of the relationship between the UK and the USA. News media in both countries cover each other more than other countries. At the same time, they are covered by many other countries, and thus, the volume of media attention from the UK to the USA is not significant compared to that which the USA received from the rest of the world (and vice versa).

The former colonial ties, which have enduring political, economic, ideological, and cultural relationships, also show the significant dependency of media attention from former colonies to the former colonial state. From the 145 colonial ties compiled by [17], after removing the relationships that do not pay or get significant attention ($[c_i \cdot \rightarrow \cdot \cdot \leftarrow \cdot c_j]$), 48.3% of them are $[c_i \triangleright \rightarrow \triangleright \cdot \leftarrow \cdot c_j]$ and 20.7% are $[c_i \triangleright \rightarrow \cdot \cdot \leftarrow \cdot c_j]$. Even after independence, 67.0% of the post-colonial ties show dependent relationships. This demonstrates that the former colonial ties are important for understanding media attention as well as migration, economic trade, and military base presence.

Furthermore, the profile of a country, which represents how it is involved in each type of relationship, can be used to classify how the country manages different relationships in the media attention network. For example, although the USA and Syria have a similar PageRank in Table II, their profiles are strikingly different. For example, there are 53 [USA $\cdot \rightarrow \cdot \cdot \leftarrow \triangleleft c_j$] cases but only 11 [Syria $\cdot \rightarrow \cdot \cdot \leftarrow \triangleleft c_j$] cases. We will leave for future work the comparison and clustering of country profiles.

TABLE V
TAXONOMY OF DISREGARD RELATIONSHIPS

| Desc | | Count | Example $[c_i, c_j]$ |
|---|---|---|---|
| $c_i \triangleright \rightarrow \triangleright$ | $\triangleleft \leftarrow \triangleleft c_j$ | - | - |
| $c_i \triangleright \rightarrow \triangleright$ | $\triangleleft \leftarrow \cdot c_j$ | - | - |
| $c_i \triangleright \rightarrow \triangleright$ | $\cdot \leftarrow \cdot c_j$ | 1 | [Turkey, UK] |
| $c_i \triangleright \rightarrow \triangleright$ | $\cdot \leftarrow \cdot c_j$ | 13 | [Japan, USA], [Vietnam, USA] |
| $c_i \triangleright \rightarrow \cdot$ | $\triangleleft \leftarrow \triangleleft c_j$ | 1 | [Israel, China] |
| $c_i \triangleright \rightarrow \cdot$ | $\cdot \leftarrow \triangleleft c_j$ | 11 | [Saudi Arabia, China], [China, Turkey] |
| $c_i \triangleright \rightarrow \cdot$ | $\cdot \leftarrow \cdot c_j$ | 478 | [Germany, Russia], [Canada, Egypt] |
| $c_i \cdot \rightarrow \triangleright$ | $\cdot \leftarrow \cdot c_j$ | 1 | [Germany, India] |
| $c_i \cdot \rightarrow \triangleright$ | $\cdot \leftarrow \cdot c_j$ | 35 | [France, South Korea], [China, Spain] |
| $c_i \cdot \rightarrow \cdot$ | $\cdot \leftarrow \cdot c_j$ | 2875 | [UK, Greece], [Pakistan, Canada] |

Table V shows the taxonomy of pairwise country relationships in $N^D$. For both countries in both directions, 84.2% of relationships are non-significant, meaning that more relationships have significant links in $N^D$ than $N^A$.

We explain each type with more than one case. First, followed by the bi-directed non-significant type, the second most frequently observed type is $[c_i \triangleright \rightarrow \cdots \leftarrow \cdot c_j]$. In this pattern, $c_i$ disregards $c_j$ significantly, and many other countries also disregard $c_j$. This is well explained by the embedded star-shaped structures in $N^D$. Secondly, 35 country relationships follow $[c_i \cdot \rightarrow \triangleright \cdot \leftarrow \cdot c_j]$. One possible explanation for this pattern is that $c_j$ is usually not disregarded by other countries and thus, even a small number of disregarded cases can cause this significance. Or, $c_i$ mainly focuses on domestic issues and disregards foreign issues. Thus, all the outgoing links have high enough weights to be significant to a receiver. Thirdly, 13 cases follow $[c_i \triangleright \rightarrow \triangleright \cdot \leftarrow \cdot c_j]$. This pattern indicates that $c_i$ frequently disregards events happening in $c_j$ even though they are newsworthy for many other countries. Lastly, $[c_i \triangleright \rightarrow \cdot \cdot \leftarrow \cdot \vartriangleleft c_j]$ shows that both countries disregard each other at a significant level, but since many other countries also disregard those two countries, the level of disregard from each other is not significant. Some relationships between hub countries follow this pattern.

In summary, we find that the media attention network is formed asymmetrically and hierarchically. We then add a new dimension of pairwise relationships: whether the media attention from or to a country is significant for a sender or receiver. This provides a richer understanding of the media attention and disregard among countries.

### E. At Triadic Level: Network Motif

In this section, we advance our understanding of the structure of the MADN by examining the network at a triadic level. In particular, network motif analysis can characterize the network regarding the information transmission mechanism. Also, extending the dyadic analysis, we further investigate the asymmetric structure of the MAND.

A network motif is a small-sized (usually 3 or 4) subgraph. Milo et al. reported that complex networks even with similar degree distributions have different profiles of motifs based on their types, such as social networks, biological networks, or machine networks [18]. For instance, social networks are more likely to have fully-connected triads than other networks due to high reciprocity of social interaction.

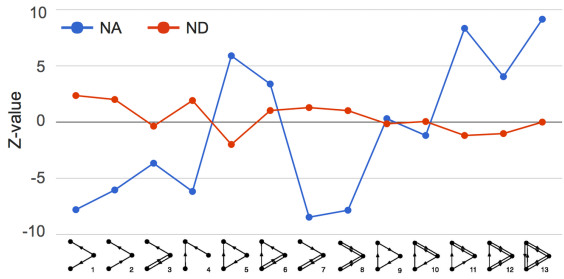

Fig. 4. Motif Profiles in $N^A$ and $N^D$

We use muxViz to generate profiles of network motifs [19]. As a null model, we generate 5,000 random networks with the same degree distribution of the original network. By comparison with this null model, we identify the significant motifs in $N^A$ and $N^D$. Figure 4 shows the profiles of the network motifs of $N^A$ and $N^D$.

Three motifs, which are motifs 5, 6, and 11, are statistically significant in $N^A$ (p<.05). The motifs 5 and 6 are called feed-forward loop (FFL) and double feedback loop, respectively. These motifs signify the existence of hierarchy in a given network. In a food web network, a primary predator eats a prey, but a secondary predator eats a prey and the primary predator as well. In this case, an FFL is formed with a prey whose incoming links are two [18]. Similarly, in a social Q&A service, a lower-level expert can answer easy questions, but a higher-level expert can answer both easy and difficult questions, and thus, an FFL is formed [20]. In media attention network, the feedforward loop indicates that media attention has a transitive hierarchy: one country pays attention to two countries, and one of the two countries pays attention to the other country in the two.

Like the motifs 5 and 6, the motif 11 also implies the existence of the country that does not get any significant attention from others but pays attention to others. While the motifs 12 and 13 have high Z-value, it is not statistically significant because the number of occurrences of those triads is only 3 and 1, respectively.

From Figure 4, the network motif profiles of $N^D$ are different from those of $N^A$, implying the network structures of $N^A$ and $N^D$ are hugely different in this scale. Three motifs, which are motif 1,2, and 4, are significant with p-value < 0.05. It is understandable that motif 1 (fan-out) is significant in $N^D$ because, as we mentioned earlier, only 10 countries have all the incoming links. Similarly, motif 4 (fan-in) is significant. These two motifs are led by star-shaped subnetworks of $N^D$. Another overrepresented motif, motif 2 (cascade), is somewhat interesting because it can be a feedforward loop with an additional link, meaning that transitivity does not hold disregard.

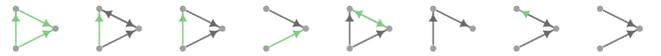

Fig. 5. 8 significant colored motifs sorted by $|Z|$

Figure 5 shows the 8 significant colored motifs (where p < .01 and $|Z|$ >1) computed by muxViz [19], where the green colored link maps into attention, and the gray colored link maps into disregard. All of them are overrepresented, and no underrepresented motifs exist in the MADN compared to the null model. We can see that FFL in $N^A$ and cascade and fan-in in $N^D$ are also significant in the MADN. The fan-out motif in $N^D$ changes to the 5th motif in Figure 5 by adding an attention exchange between two countries that were disregarded. Also, the fan-in motif in $N^D$ forms variations (the 3rd, 7th, and 8th motif in Figure 5) by adding attention relationships. The colored motif profile stresses that the MADN is similar to the sum of the separate networks but also different enough to have microstructures (triads) of attention and disregard together that cannot be observed from the attention or disregard networks.

### F. At Community Level: Global Village and Unique Blocks

Lastly, we conduct a community analysis to capture the structural characteristics of relationships among more than three countries. We find eleven communities in $N^A$ by the InfoMAP method [21]. We use the backbone network because the whole network is too dense to find any modular structure besides one huge community having 114 countries.

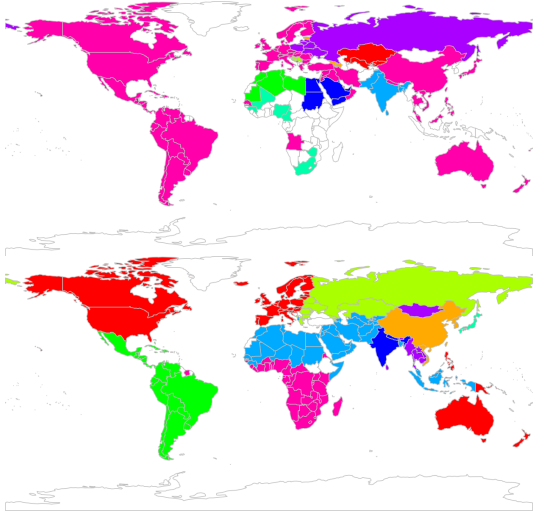

Fig. 6. (Top) Communities of countries exchanging media attention, (Bottom) Clash of Civilization by Huntington [22]

Figure 6 [top] depicts the world map grouped by eleven communities of media attention. The identified communities are to some extent aligned with the concept of "civilizations" proposed by Huntington, which groups the world by culture and religious identity [22]. Recently, the persistence of civilizations has been reported from large-scale social datasets, such as e-mail and Twitter communications [23].

Among eleven communities, the largest community contains 83 countries around the world, implying the globalization of the media attention. For these countries, media attention crosses cultural boundaries and moves freely, resulting in one giant community of media attention.

The second-largest community consists of Russia and its neighbors: Lithuania, Ukraine, Russia, Belarus, Poland, and Latvia, and the third largest community consists of the seven MENA countries: Bahrain, Saudi Arabia, Yemen, the UAE, Kuwait, Sudan, and Egypt. Interestingly, Qatar, which has Al Jazeera, is not included in this cluster. The news media in these countries frequently cover each other, which makes sense as not only they are physically close together, but also they are close culturally, economically, and politically. The members of the two communities are subsets of "Orthodox" and "Islamic" civilizations [22], respectively.

The above examples imply that those small communities can be created by physical, cultural, economic, and political closeness. In our data, we find other small communities that can be explained this way. One community consists of four countries in Southern Asia: Pakistan, India, Bangladesh, and Sri Lanka, which are aligned with the "Muslim" civilization and "Cleft" countries [22]. The six South African countries (Cameroon, Guinea, Mali, South Africa, Nigeria, Zimbabwe), and the five North African countries (Libya, Algeria, Morocco, Tunisia, and Mauritania), form a community each. Lastly, we find a community comprised of the three Southern Europe countries: Bosnia and Herzegovina, Croatia, Serbia. They are not only close but also share a historical background–they used to be a part of Yugoslavia. As each of the remaining communities consists of two adjacent countries only, we omit them in our explain.

While the news world becomes a global village, we still see small communities, generally aligning with the concept of civilizations proposed by Huntington. The African communities (Islamic+African) and the southern European community (Orthodox+Islamic) cross civilizations, but they are geographically close or share a historical background.

In $N^D$, interestingly, we find only one community for the whole world, which means that there is no group of countries that is disregarded more than others.

### G. At Network Level: Vector Representation of MADN

The "global village" trend is also found in the vector representation of the nodes learned by Node2Vec [24] in $N^A$. In the resulting vector space, the vectors show neighboring relationships by calculus, such as v(Australia) - v(New Zealand) + v(Tunisia) = v(Algeria), or v(Qatar) - v(UAE) + v(Malaysia) = v(Philippines), proving that the learning of the node representation is well done. We visualize the vectors of countries by applying t-SNE, a dimensionality reduction method based on manifold learning [25] in Figure 7. The color of a circle represents the sub-region of the world.

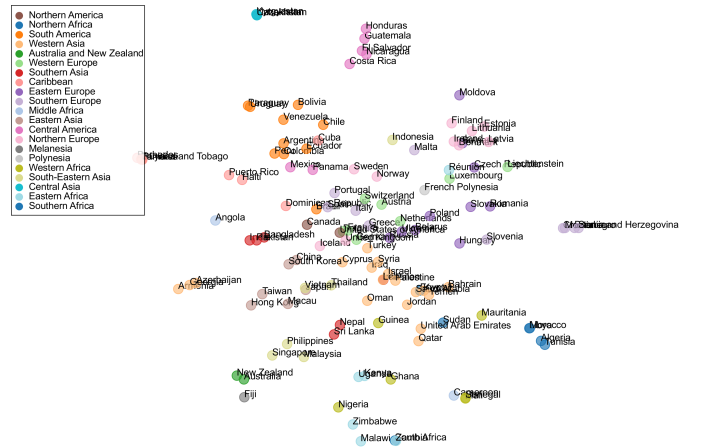

Fig. 7. Node2Vec in $N^A$

In the figure, we see the same colored circles (countries in the same region) are located closely, indicating they have similar media attention patterns. We find previously observed Russian (red), MENA (orange), African (navy), and other communities. Two or three countries that are closely located, such as [Singapore, Indonesia, Philippine], show that their media attention is more similar to each other, while they are members of a big community. At the center of the figure, the countries of the different regions are not well divided, even though they are surrounded by the countries of the same region. This pattern is consistent with the "global village" trend and shows local uniqueness.

The representations of countries in $N^D$ are completely aligned with the community analysis – countries of the same region are dispersed all over the vector space.

## V. CONCLUSION

In this work, we investigate the structural characteristics of the multiplex media attention and disregard network (MADN) among 129 countries. This first approach to the topology of the MADN introduces a new way to understand the international order of media attention. From the multi-level analysis, we see the skewed, hierarchical, and asymmetric structure of the MADN. Also, we observe that the media attention follows the "global village" trend, but at the same time, unique attention blocks of the MENA region, as well as Russia and its neighbors, still exist.

One limitation of this work is that we use one source for the analysis. While Unfiltered News uses one of the largest, most reliable, and least biased news aggregating systems, i.e., Google News with 75K news sources, the algorithms they used to process the data are still unknown. It is a general limitation in studying an external service. Nevertheless, Unfiltered News has been actively studied recently and shows reliability in estimating the press freedom index based on news coverage [26] or empirical evidence in the alarm/patrol hybrid model of media attention [27]. Also, unlike other global news datasets, such as GDELT or EventRegistry, Google News has better quality control since it is a service for millions of visitors. Thus, we believe that studying Unfiltered News adds a valuable asset to the data-driven study of journalism. The other potential limitation is that the amount of press coverage does not consider the context–why a certain country is reported by others. Nevertheless, previous studies, based on the volume of foreign news coverage, have discovered international relations among countries without consideration of context, and, additionally, our work takes "disregard" into consideration to mitigate this issue.

For the future direction, we would dig into the comparison with other networks to represent country relationships, such as migration, trade, or flight networks, and reveal their interplay.